\documentstyle[pre,aps]{revtex}
\begin{document}
\title{Spatial-temporal correlations in the process to self-organized
criticality}
\author{C.B. Yang$^1$, X. Cai$^1$ and Z.M. Zhou$^2$}
\address{
$^1$\ \ Institute of Particle Physics, Hua-Zhong 
Normal University, Wuhan 430079, China}
\address{$^2$\ \ Physics Department, Hua-Zhong University of
Science and Technology, Wuhan 430074, China}
\date{\today}
\maketitle

\begin{abstract}
A new type of spatial-temporal correlation in the process approaching 
to the self-organized criticality is investigated for the two simple
models for biological evolution. The change behaviors of the position
with minimum barrier are shown to be quantitatively different in the
two models. Different results of the correlation are given for the
two models. We argue that the correlation can be used,
together with the power-law distributions, as criteria for
self-organized criticality.

\end{abstract}

The phenomenon of ``self-organized criticality'' (SOC), with potential
applications ranging from the behavior of sandpile and the description
of the growth of surfaces to generic description of biological evolution,
has become as a topic of considerable interest
\cite{btw87,cbo91,bcc89,s92,bc91,cb89,bcc90,ss89}.
It is observed that the dynamics of complex
systems in nature does not follow a smooth, gradual path, instead it
often occurs in terms of punctuations, or ``avalanches'' in other word.
The appearance of the spatial-temporal complexity in nature, containing
information over a wide range of length and time scale, presents a
fascinating but longstanding puzzle.  Such complexity also shows up
in simple mathematical models for biological evolution and growth
phenomena far from equilibrium. In former studies, power-law
distributions for the spatial size and lifetime of the ``avalanches''
have been observed in various complex systems and are regarded
as ``fingerprints'' for SOC. It seems that there is no general
agreement on a suitable definition of SOC \cite{s94,ccc94}, although
a minimal definition was given in \cite{f96}.
Because there is no universally accepted ``black-box''
tests for the presence or absence of SOC based solely on observables, 
systems with a wide range of characteristics have all been designated
as ``self-organized critical''.

While numerous numerical studies have claimed SOC to occur in specific
models, and although the transition to the SOC state was studied in
\cite{pmb94,pmb96,corr99}, a question has never been answered:
How is the process approaching to the final dynamical SOC attractor
characterized? One may even ask whether the phenomenon SOC can be
adequately characterized by such power-law distributions. The answer
to the latter question seems to be negative, as concluded in
\cite{bjw95}. In Ref.\cite{bjw95} were pointed out ``some striking 
observable differences between two `self-organized critical' models which
have a remarkable structural similarity''. The two models, as called
the Bak-Sneppen (B-S) models, are introduced in 
\cite{bs93,fbs93,bdfjw94} and are used to mimic biological evolution.
The models involve a one-dimensional random array on $L$ sites. Each
site represents a species in the ``food-chain''. The random number
(or barrier) assigned to each site is a measure of the ``survivability''
of the species. Initially, the random number for each species
is drawn uniformly from the interval (0, 1). In each update, the least
survivable species (the update center) and some others undergo mutations
and obtain new random numbers which are also drawn uniformly from (0, 1).
In the first version of the model (the local or nearest-neighbor model),
only the update center and its two nearest neighbors 
participate the mutations. In the second version, $K-1$ other
sites chosen randomly besides the update center are involved in the
update and assigned new random survivabilities (so this version is
called random neighbor model). Periodic boundary conditions are adopted
in the first model. As shown in \cite{bdfjw94,jpa95,ymp97}, the second 
version is analytically solvable. Investigation in \cite{bjw95} shows
that some behaviors of the local and random neighbor models are qualitatively
identical. They both have a nontrivial distribution of barrier heights of
minimum barriers, and each has a power-law avalanche distribution. But the
spatial and temporal correlations between the minimum barriers show different
behaviors in the two models and thus can be used to distinguish them.

In all the studies mentioned above, spatial and/or temporal distributions of
the ``avalanches'' and correlations between positions with minimum of 
barriers are investigated separately. As shown in many studies, however,
spatial and/or temporal distribution of the ``avalanches'' alone cannot be
used as a criterion for SOC, nor can the spatial or temporal correlation
do. In this paper, it is attempted to study a new kind of
correlation between minimum barriers in the process of the updating in the
two models for biological evolution. The correlation between the positions
with minimum barriers at time (or update) $s$ and $s+1$ is investigated.
Since the new correlation involves {\em two sites} at {\em different} times,
it is of spatial-temporal type. Thus it may be suitable for the study of 
spatial-temporal complexity.

Consider the update process of the local neighbor model. Initially,
each site is assigned a random number. All the random numbers are drawn
uniformly from interval (0,1). Denote $X(s)$ the site number with minimum
barrier after $s$ updates. The sites can be numbered such that $1\le
X(s)\le L$. To see how $X(s)$ changes in updating process in the model
$X(s)$ is shown in Fig. 1 as a function of $s$ for an arbitrary update
process for lattice size $L$=200 with $s$ from 1 to 2000. The lower part
of Fig. 1 is a zoomed part of the upper one for small $s$.
It is clear that $X(s)$ seems to be random when $s$ is small. With
the going-on of updating, $X(s)$ becomes more and more likely to be
in the neighborhood of last update center, $X(s-1)$. So there appear
some plateau like parts in Fig. 1. In other word,
there appears some correlation between $X(s)$ when the system is 
self-organized to approach the critical state. So, it may be fruitful to
study the self-correlation of $X(s)$ in searching quantities characterizing
the process to SOC. For this purpose, one can
define a quantity
\begin{equation}
C(s)=\langle X(s)X(s+1)\rangle-\langle X(s)\rangle\langle
X(s+1)\rangle \quad ,
\end{equation}

\noindent with average over different events of updating. Obviously,
if there is no correlation between the sites with minimum barrier at time
$s$ and $s+1$, or $\langle X(s)X(s+1)\rangle=\langle X(s)\rangle
\langle X(s+1)\rangle$, $C(s)$ will be zero. Thus, $C(s)$
can show whether there is correlation between $X(s)$ and also give a
measure of the strength of the correlation. Because of the randomness of
the survivability at each site, $X(s)$ can be 1, 2, $\cdots$, $L$ with
equal probability, $1/L$. Thus, $\langle X(s)
\rangle=(L+1)/2$ for every time $s$. It should be pointed out that
$\langle X(s)\rangle=(L+1)/2$ does not mean any privilege of sites with
numbering about (L+1)/2. In fact, all sites can be the update center with
equal chance at time $s$ if the update process is repeated many times from the
initial state. Due to the randomness of the updated survivability $X(s+1)$
can also take any integer from 1 to $L$. However, the distribution of
$X(s+1)$ is peaked at $X(s)$ when $s$ is large, see \cite{pmb96} for detail.
With the update going on, the width of the distribution becomes more and
more narrower. When the width becomes narrow enough, $\langle X(s)X(s+1)
\rangle$ will turn out to be $\langle X^2(s)\rangle=(2L^2+3L+1)/6$. So,
$C(s)$ will approach $(L^2-1)/12$ for large $s$.
In above definition
for $C(s)$, however, the neighboring relation  between $X(s)$ and $X(s+1)$
cannot be realized once the numbering for the sites is given.
Due to the periodic boundary conditions adopted in the model,
one of the nearest neighbors of the site with numbering 1
is the one numbered $L$. To overcome this shortcoming, one can
introduce an {\em orientational shorter distance} $\Delta(s)$ between
$X(s)$ and $X(s+1)$. Imagine the $L$ sites with numbering $1, 2, \cdots, L$
are placed on a circle in clockwise order. Then $|\Delta(s)|$
is the shorter distance between the two sites on the circle.
If $X(s+1)$ is reached along the shorter curve from $X(s)$
in clockwise direction, $\Delta(s)$ is positive. Otherwise $\Delta(s)$
is negative. For definiteness, one can assume $-L/2\leq \Delta(s)< L/2$.
With $\Delta(s)$, one can use 
\begin{equation}
X^\prime(s+1)=X(s)+\Delta(s)
\end{equation}

\noindent
in place of $X(s+1)$ in the definition of $C(s)$. Since $X^\prime(s)$
can cross the (non-existing) boundary between 1 and $L$ and reflect the
neighboring relation with $X(s)$, the effect of periodic
boundary conditions on the correlation can be taken into account.
(In the simulation of the B-S model numbering the $L$ sites with
integer numbers $1, 2, \cdots, L$ is necessary, but the start position
can be arbitrary. Different numbering scheme will give the same results for
$C(s)$, as physically demanded. This in return is also an indication 
of the equivalence of all sites in the presence of periodic boundary
conditions.)
To normalize the dependence of $C(s)$ on the size of the
one-dimensional array, we can renormalize $C(s)$ by $(L^2-1)/12$.
In the following, we use a normalized definition of $C(s)$ as
\begin{equation}
C(s)={\langle X(s)X^\prime(s+1)\rangle-\langle X(s)\rangle\langle X^\prime(s+1)
\rangle\over (L^2-1)/12}\ .
\end{equation}

In current study $X(s)$ and $\Delta(s)$ are determined from Monte Carlo
simulations, and 500,000 simulation events 
are used to determine the averages involved. For each event, 
2000 updates are performed from an initial state with random barriers on the
sites uniformly distributed in (0, 1). The normalized correlation function
$C(s)$ is shown as a function of $s$ in Fig. 2 for $L=50,\ \ 100,$ and
200. One can see that $C(s)$ is a monotonously increasing function of
time $s$. As in our naive consideration, $C(s)$ is very small in
the early stage of updates and becomes larger and larger for larger
$s$, indicating the increase of the strength of correlation between
the sites with minimum barrier at different times. The behavior of
$C(s)$ with $s$ exhibits different characteristics for small and large
$s$. $C(s)$ increases with $s$ very quickly for small $s$, but the
rate becomes quite slow after a knee point. The knee point appears
earlier for smaller $L$, showing the existence of a finite-size effect. 
Also, the seemly saturating value of $C(s)$
depends on the size $L$ of the lattice, or more clearly, it increases
with the lattice size $L$. Since only 500,000 simulation events
are used in current study, there shows the effect of fluctuations
in the figure.

The correlation between $X(s)$ can be investigated for the random
neighbor model for biological evolution in the same way. For simplicity
only the case with $K=3$ is taken into account. The generalization to
other cases is straight forward. First,
one can have a look on how $X(s)$ changes with update. $X(s)$ is shown
as a function of $s$ in the upper part of Fig. 3.
This plot may look as a random scatter of points at first sight.
But it is not. A close look reveals correlations: $X(s)$ often has
almost same value for several consecutive or almost consecutive
$s$ values.  However, no obvious plateau like part can be seen in the
figure, showing the difference between the two versions of B-S model.
$C(s)$ is also studied and shown in the lower part of Fig. 3 as
a function of $s$ for the lattice size $L=200$. In the random neighbor
version of the B-S model, sites numbered with 1 and $L$ are no longer
neighbors. So, in the calculation of $C(s)$ from Eq. (3), $X(s+1)$
is used instead of $X^\prime(s+1)$. The counterpart
for the nearest neighbor model is also drawn in the figure for
comparison. One can see that the saturating value is much smaller
than in the case of the local neighbor version of the model. 

From the discussions above one can see that the correlation between
the sites with minimum barrier may play an important role in 
investigating SOC. The power-law distributions for the size and 
lifetime of the ``avalanches'' together with the new kind of correlation 
may be used as criteria for SOC.
 
This work was supported in part by the NNSF in China and NSF in Hubei, China.
One of the authors (C.B.Yang) would like to thank Alexander von Humboldt
Foundation of Germany for the Research Fellowship granted to him.

\begin{center}
{\Large\bf Figure Captions}
\end{center}

\begin{description}
\item{Fig. 1} The change of site $X(s)$  
with time $s$ for an arbitrary event in the nearest neighbor version
of the B-S model for biological evolution.

\item{Fig.2} The correlation function $C(s)$ as a function
of $s$ for lattice size $L$=50, 100, and 200 for the same model
as in Fig. 1.

\item{Fig. 3} Upper part: The change of site $X(s)$  with $s$ for
the random neighbor version of the B-S model for biological evolution;
Lower part: The correlation function $C(s)$ for the two versions 
as functions of $s$ for $L$=200.

\end{description}
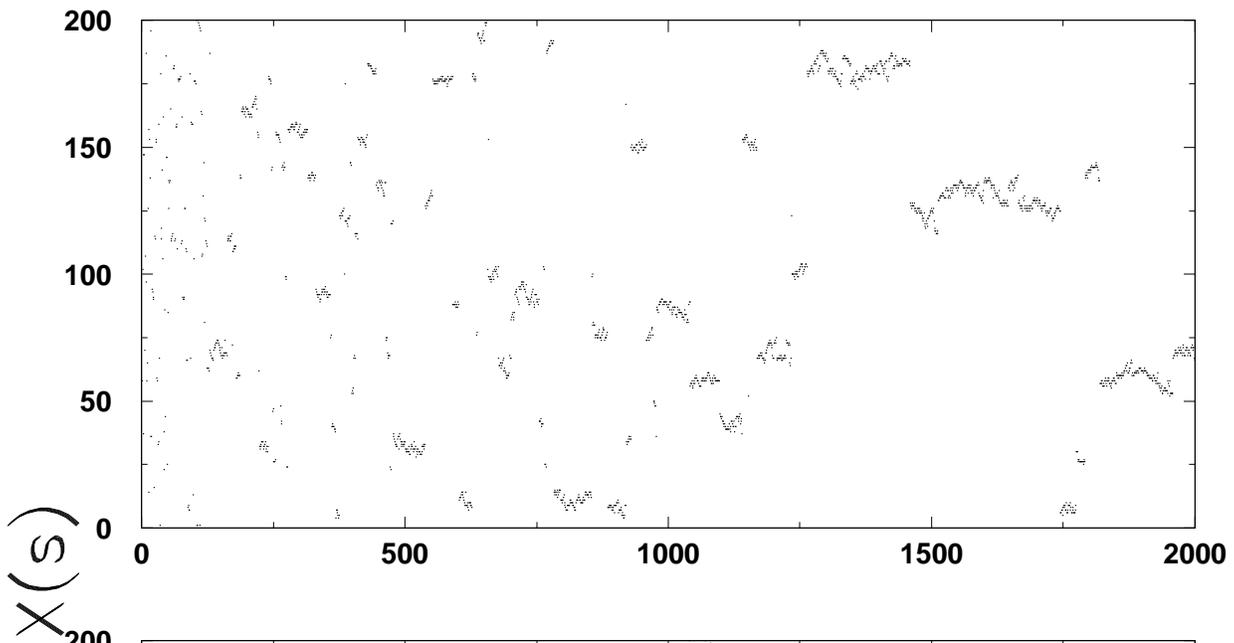
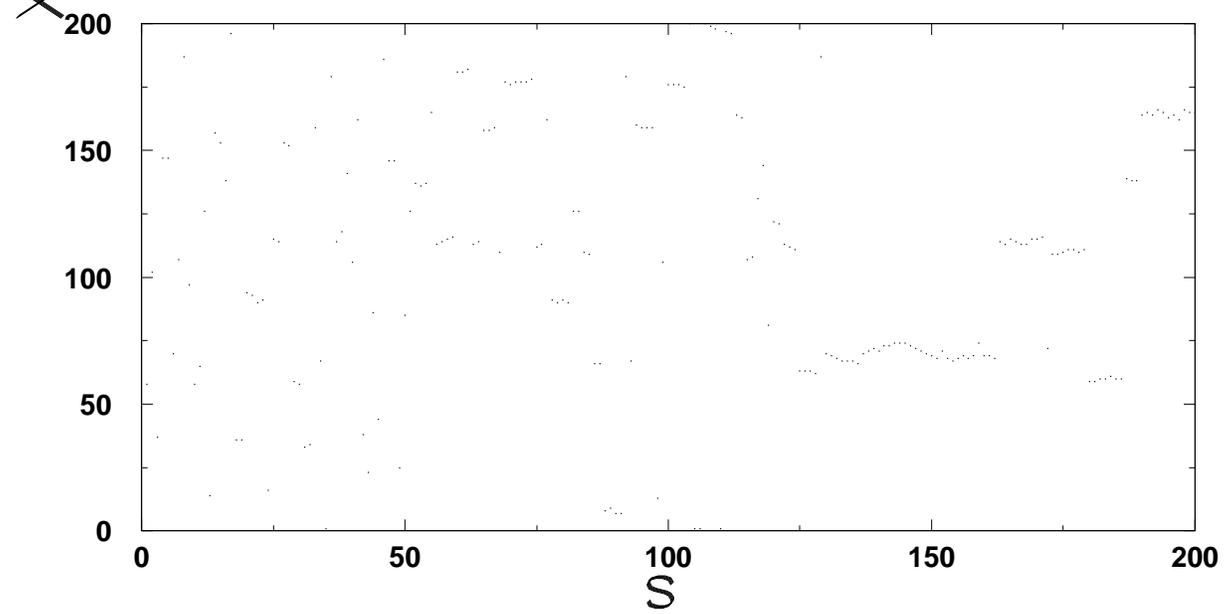
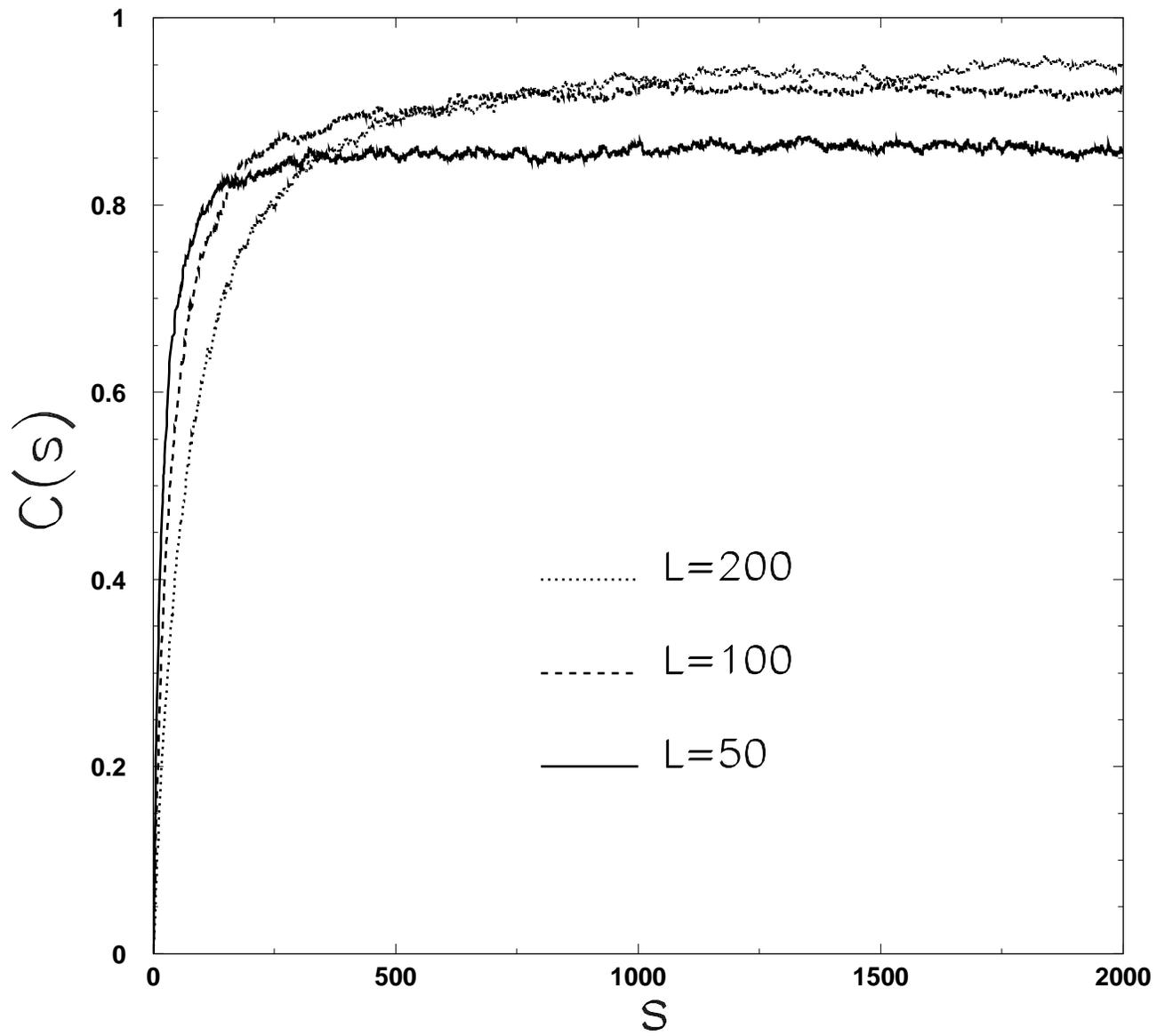
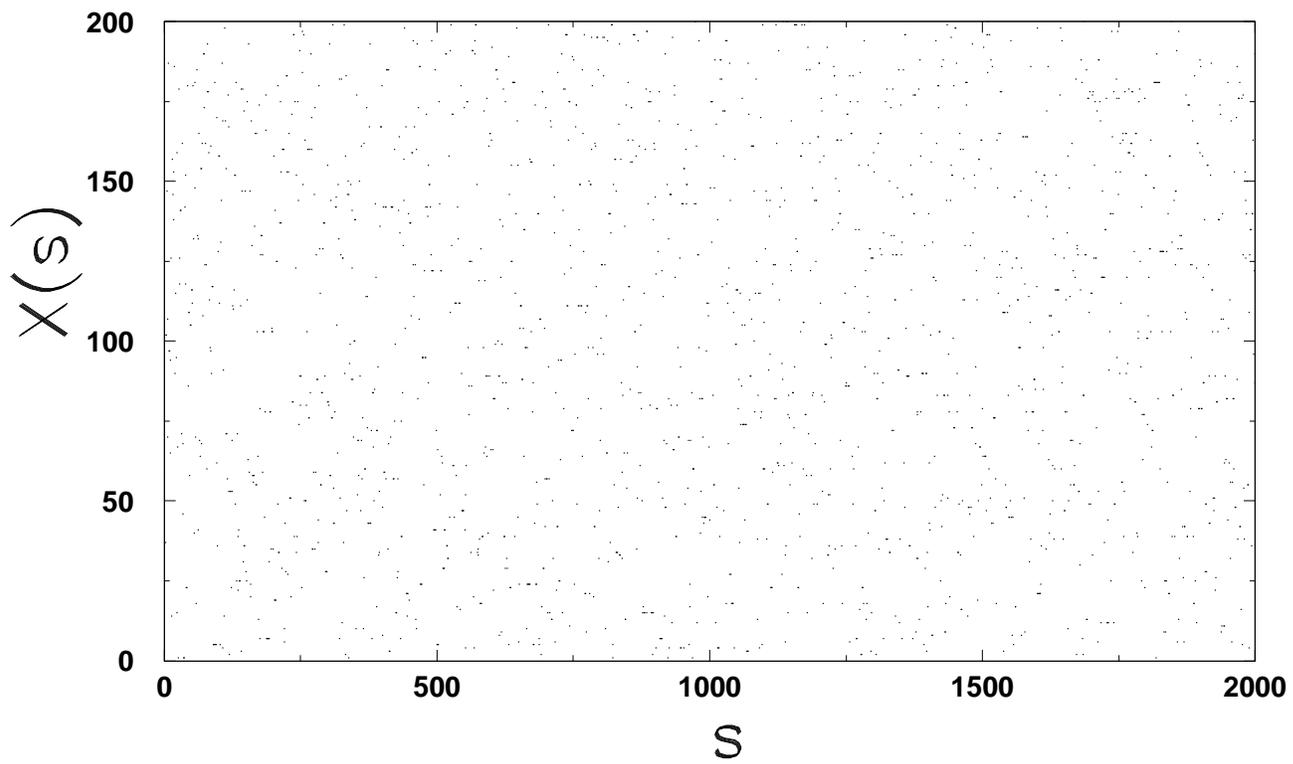
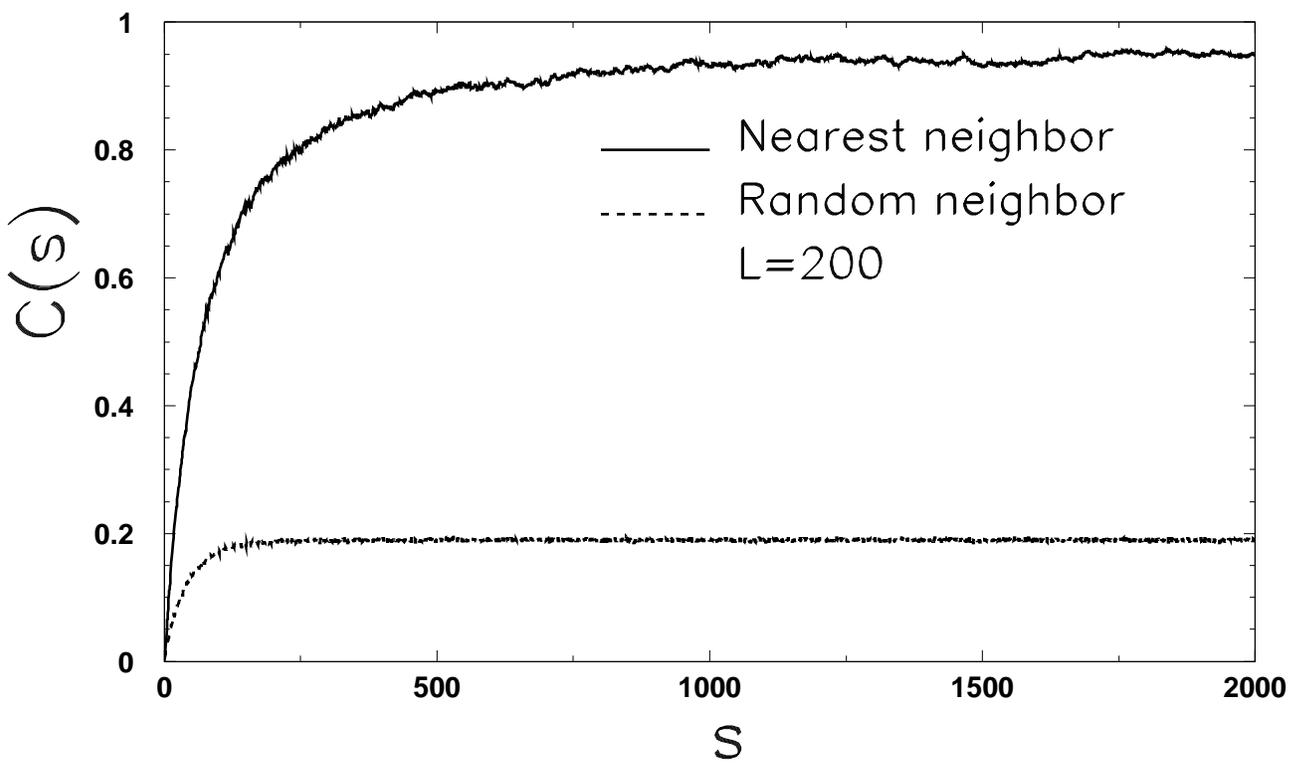
\end{document}